\documentclass[11pt,a4paper]{article}

\usepackage{amssymb}
\usepackage{amsmath}

\newcommand{\pa}{\partial}
\newcommand{\Rset}{{\mathbb R}}
\newcommand{\Cset}{{\mathbb C}}

\newcommand{\myref}[1]{(\ref{#1})}

\newcommand{\om}{\omega} 

\newcommand{\de}{\delta}
\newcommand{\al}{\alpha}
\newcommand{\eps}{\epsilon}
\newcommand{\la}{\lambda}
\newcommand{\ga}{\gamma}
\newcommand{\Ga}{\Gamma}
\newcommand{\be}{\beta}
\newcommand{\te}{\theta}

\renewcommand{\leq}{\leqslant}
\renewcommand{\geq}{\geqslant}
\newcommand{\lan}{\langle}
\newcommand{\ran}{\rangle}

\newcommand{\sur}[2]{{\displaystyle\mathop{#1}_{#2}}}

\newlength{\somme}
\newlength{\sommep}
\newcommand{\intvide}{\rule[-\sommep]{0cm}{\somme}}

\newlength{\sommebis}
\newlength{\sommepbis}
\newcommand{\demi}{\frac{1}{2}}

\usepackage{amssymb}
\usepackage{amsmath}
\usepackage{epsfig}
\usepackage{latexsym}
\usepackage{color}
\usepackage{anysize}

\marginsize{2cm}{2cm}{2cm}{2cm}

\renewcommand{\th}{\theta}

\newcommand{\tla}{\widetilde{\la}}
\newcommand{\teps}{\tilde{\varepsilon}}

\renewcommand{\eps}{\varepsilon}
\renewcommand{\geq}{\geqslant}
\renewcommand{\leq}{\leqslant}

\newcommand{\wz}{\mathbf{w_0}}

\newcommand{\tz}{\tau_0}
\newcommand{\hga}{\widehat{\ga}}
\newcommand{\htz}{\widehat{\tau}_0}
\newcommand{\hal}{\widehat{\al}}
\newcommand{\ho}{\widehat{\omega}}
\newcommand{\hbe}{\widehat{\beta}}
\newcommand{\piz}{\left\langle e^{-\la\tau\eps}\right\rangle_\wz}
\newcommand{\bareps}{\overline{\eps}}

\begin{document}
\title{Power fluctuations in stochastic models of dissipative systems.}
\author{Jean Farago\footnote{e-mail: farago@lcp.u-psud.fr. Tel: (33)-1-69-15-41-95}} 
\date{\small \it LCP-UMR 8000, Bat. 349, Universit\'e Paris-Sud, 91405 Orsay cedex, France }
\maketitle

\abstract{We consider different models of stochastic dissipative equations and
  theoretically compute  the probability distribution functions
  (actually the associated large deviation functions) of the
  \textit{time averaged} injected power required to sustain a
  nontrivial stationary state. We discuss the results and in
  particular draw from our results some general features shared by
  these distributions in realistic dissipative systems.}

\medskip

\noindent PACS: 05.40.-a: Fluctuation phenomena, random processes, noise, and Brownian motion.\\
\phantom{PACS:} 02.50.-r: Probability theory, stochastic processes, and statistics.

\section{Introduction}

Among out of equilibrium statistical systems, strongly dissipative
mediums have a very particular status. Usually, statistical
theories of non equilibrium systems deal with tools inspired or
inherited from thermal equilibrium, a scheme that  Onsager theory
examplifies especially well \cite{kubo,degroot}. That such a continuity may be possible in
numerous cases is
always due to the fact that  genuine equilibrium
keeps a certain relevance, thanks to a time decoupling which induces partial
equilibrations \cite{grabert}. Even in recent theories of
nonequilibrium statistical mechanics aiming at describing systems as
complicated as glassy materials, equilibrium concepts are generalized
and successfully adapted, since again time
separation is there at work \cite{kurchanCRAS}.

In contrast, strongly dissipative systems have a full rest natural ``fixed point'' or equilibrium  (something like
a zero temperature situation). To avoid such a  state
and explore the richness of their dynamics, the system has to be fed continuously with energy, by means of an external action which
holds it in a non trivial stationary state (if the injection mechanism
has itself some stationarity properties). As a result, the statistical
state (i.e. the stationary measure) reached is very far from any concept of (Boltzmann-like)
equilibrium, and new approaches must be followed. For instance, granular matter
 looks more or less like a gas when sufficiently shaked; actually, the
 so-called ``granular temperature'' defined as the mean kinetic energy
 per particle, is a rather hazy notion, since its values are not
 unique  when the system is not monodisperse \cite{wildman,feitosa}, in a clear
 violation of  a
 basic requirement for a temperature-- that is, the equilibration
 of temperatures of any pair of subsystems at equilibrium (or
 equivalently, the energy equipartition principle).

Therefore, these systems, sometimes termed ``far from equilibrium'',
do not belong to the traditional statistical physics (as far as we are
concerned with  relevant degrees of freedom: of course, the Navier-Stokes
equation can be derived from local thermodynamical
equilibrium considerations; however, a turbulent velocity field obeys a
dissipative dynamical equation (NS, say) and its statistics  has nothing to do with standard statistical physics) and
consequently requires alternative approaches. 

 Some years ago, 
 such an original
approach was proposed \cite{labbe}, which  tried to
look at the problem of dissipative systems from a \textit{global} and phenomenological
point of view. In particular, it was recognized that the common
feature of all dissipative systems is the energy evolution equation
$\dot{E}=I-D$, which explicites and summarizes  the
physics of these systems from a ``macroscopical'' (or experimental)
point of view : two channels of energy flow compete, one
corresponding to the energy injected by the external operator ($I$),
the other related to an inner dissipation term ($D$, easily recognized,
since it is in general a volume term proportional to a dissipation
coefficient). Of course, the physics is actually more complicated, since
these energy flows induce complicated structurations of the local
fields, which themselves determine \textit{in fine} the statistical
properties of $I$ and $D$; but it was thought that if something common
to all dissipative systems
could be once firmly asserted, this will be done necessarily using that kind of
general evolution equation, which transcends all peculiar
details of each experimental situation.

Experimental measurements were first performed in a turbulent
von Karm\'an flow, and statistical properties of $I$ were extracted
and analysed in stationary regimes. Surprisingly enough, it was shown
that the fluctuations of $I$ are non gaussian (despite its \textit{a
  priori} ``extensive'' (on the surface) nature), and  decay
abnormally slowly as Reynolds number increases (i.e. more slowly that
a normal regression
$\sim 1/\text{Re}^{1/2}$).  The analysis of these
experiments is still the subject of active research and debates \cite{bhp}, and the physical 
interpretation of the observed scaling is up to now not 
fully completed. Similar studies were afterwards made on various systems
including granular gases, shell models of turbulence, self-organized
critical systems, and led to  interesting developments: for instance, a very convincing way to
estimate the number of effective degrees of freedom in a shaked
granular gas has been proposed \cite{aumaitreepjb}. 

From a theoretical point of view, exploring  the properties of
$I$ and $D$ is not easy, for the systems considered are characterized
by invariant measures generally unknown. In a preceding paper
\cite{jean}, we studied a simple model (``zero-dimensional'') of
dissipative system, where the statistical properties of $I$ and $D$
were calculable. More precisely, we considered the following
stochastic equation
\begin{equation}
  \ddot{x}+\ga \dot{x}+V'(x) =\psi(t)
\end{equation}
where $\psi$ is a white noise and $V(x)$ a potential. We read this equation as the dynamical
evolution of a single dissipative coordinate ($\ddot{x}+\ga \dot{x}+V'(x)=0$)
subjected to an external forcing $\psi(t)$  (the stochasticity of the forcing mimics the chaoticity developped by realistic dissipative systems when driven  vigorously enough out of equilibrium; this {\it ad hoc} choice allows exact computations; we verified that purely chaotic dissipative systems (like a periodically forced asynchronous pendulum) gives qualitatively the same results, as far as time-averaged observables are concerned (see below); however, some interesting studies and comparisons could  be performed with deterministic chaotic systems).

 The energy of the
coordinate is naturally defined as $E=\dot{x}^2/2+V(x)$, and is conserved in the
absence of noise and dissipation. This energy obeys the evolution
equation
\begin{equation}
  \dot{E}=\psi \dot{x}-\ga \dot{x}^2
\end{equation}
where the above-mentioned structure $\dot{E}=I-D$ is clearly apparent. The
properties of instantaneous quantities of such systems are not
complicated to compute, but we were interested in calculating the
distribution of partially time-averaged quantities like
\begin{equation}
  \eps=\frac{1}{\tau}\int_t^{t+\tau}dt' I(t')
\end{equation}
Initially, the focus on these time-averaged variables was motivated by
the so-called Fluctuation Theorem \cite{ecm,gc,kurchan}, but beyond
this particular debate, this partial time averaging presents a major
advantage for analysis: it smoothes the short time scale dynamics, and
gives the possibility to highlight phenomena occuring at slow time
scales, broadening the analysis based only upon non dynamical
considerations. As an example, the systems considered in \cite{jean}
display a great variety of different dynamics, according to the form
of the potential $V(x)$: confined, unconfined, activated, etc\ldots
Despite these great qualitative discrepancies, we showed that the
probability density function of $\eps$ is asymptotically \textit{the
  same} in the limit of large $\tau$ (to be precise: $\lim
\frac{1}{\tau}\log P(\eps)=f(\eps)$ with $f$ independent of $V$; this
$f$ is called the large deviation function associated to $\eps$). In
this result can be seen that peculiar details of the short time dynamics
of energy have faded away through the averaging.
 We also remarked other
interesting facts concerning the pdf of injected and dissipated power
(averaged during a time $\tau$):  the  \textit{large deviation function}
 of the
injected power is curiously dependent on the initial conditions set of the system at the beginning of the averaging window: if the distribution of $\eps$ is measured, starting always with the \textit{same} status for the system (referenced here with the initial conditions of position and velocity $(x_0,v_0)$),  the \textit{large deviation function}
 of the
injected power is the same as the dissipated power; conversely, in the
permanent regime, that is when measurements of $\eps$ are performed  the (statistical) stationarity\footnote{characterized by time-independent probability distributions}  being reached, rare energy fluctuations cause a
\textit{singularity} to arise in the large deviation function of the injected power (but
not in the dissipation). 

These unexpected properties of the statistics of $\eps$ rise
immediately the question of their generality.
In the models we studied in \cite{jean}, we chose for $\psi(t)$ a Gaussian
white noise (i.e. $\lan \psi(t)\psi(t')\ran=\Gamma\de(t-t')$),  to
make the computations as simplest as possible. The purpose of this
paper is to study generalizations of the systems previously considered, and
discuss whether the
properties of $\eps$ yield new  concepts, possibly useful
for more realistic systems.

In a first part, we recall the results obtained in \cite{jean}. Then
we present some generalizations of the simple model (coloured
noise, non linear friction), and discuss the
properties of the large deviation functions of $\eps$ associated with
them. The details of the calculations are entirely postponed in appendices.

\section{Injected and dissipated power in Langevin equation}

In \cite{jean}, we studied the system
\begin{align}
\ddot{x}+\ga \dot{x}+V'(x)&=\psi(t)\\
\langle \psi(t)\psi(t')\rangle&=2\Gamma\de(t-t')
\end{align}
which was thought as an intrinsic dissipative system (driven by
$\ddot{x}+\ga \dot{x}+V'(x)=0$) driven away from equilibrium by a
Gaussian white noise. Of course, it is not the common interpretation
of a Langevin equation, which describes usually the thermalization of
a particle in a fluid; in that context $\ga\dot{x}$ and $\psi$ are two
different faces of the same action. There we abandoned that reference,
and in particular made no citation of the Einstein relation $\Gamma=\ga
k_BT$. We studied the statistical properties of
\begin{align}
  \eps=\frac{1}{\tau}\int_{0}^{\tau}dt'\dot{x}(t')\psi(t')
\end{align}
the injected power averaged over a finite time interval of length
$\tau$. This task is \textit{a priori} complicated, for $\eps$
involves explicitely the dynamics. Nevertheless we succeeded and
computed the distribution $Prob(\eps)$ for any value of $\tau$ and
\textit{any potential} (in fact any potential non repulsive  at
$\infty$). The distributions are of course potential dependent, but an
prominent feature emerged in the limit of large $\tau$. We know that
at large $\tau$, the distribution obeys the large deviation theorem:
\begin{align}
  \exists f, \ \ \lim_{\tau\rightarrow\infty}\frac{1}{\tau}\log P(\eps)=f(\eps)
\end{align}
$f$ is called the large deviation function associated with
$\eps$. This property is sometimes uncorrectly noted $P(\eps)\sim\exp(
\tau f(\eps))$. We demonstrated that this large deviation function is
in fact \textit{independent} of the potential $V(x)$, what is rather
unexpected, since the underlying dynamics is on the contrary very sensitive 
to the form of the potential. We found that if the distribution of the
initial energy (i.e. at $t=0$) is bounded, the large deviation function has the
expression
\begin{align}\label{ldelgv}
  f(\eps)=-\frac{\ga}{4\teps}(\teps-1)^2\ \ \ \text{where $\teps=\eps/\Gamma$}
\end{align}
 (the maximum of this function corresponds to $\lan \eps\ran$ and is always $\Gamma$ in this model since $\lan I\ran=\lan D\ran=\ga\lan \dot{x}^2\ran$). Interestingly enough, this expression is no longer valid if the
distribution of initial energy is unbounded, which is the case in the
permanent regime since $Prob(v^2)\propto \exp(-\Gamma v^2/2\ga)/|v|$. In
that case, a phenomenon analogous to a phase transition occurs and
\myref{ldelgv} is valid only above $\teps_c=1/3$. Below that value,
the curve is replaced by a portion of a straight line:
\begin{align}
  f(\eps)=\left\{
  \begin{array}{ll}\displaystyle
-\frac{\ga}{4\teps}(\teps-1)^2 & \text{if $\teps\geq 1/3$}\\
-\ga(1-2\teps) & \text{if $\teps\leq 1/3$}
  \end{array}\right.
\end{align}
(A physical explanation of this singularity is given in
\cite{jean}). Finally we computed the large deviation associated with the
dissipation $\de=\frac{1}{\tau}\int_0^\tau dt \ga \dot{x}^2(t)$ and
found that it is equal to \myref{ldelgv}, $\tilde{\de}=\de/\Gamma$ replacing $\teps$.
\section{Free and harmonically bounded particle driven by coloured noise}\label{coloured}

We now consider a slightly different system. The dynamics is given by
\begin{align}\label{eqbase}
\ddot{x}+\ga \dot{x}+\om^2x&=\psi(t)\\
\langle \psi(t)\psi(t')\rangle&=\frac{\Gamma}{\tz}\exp\left(-\frac{|t-t'|}{\tz}\right)\nonumber
\end{align}
It resembles the previous case (with $V(x)=\om^2x^2/2$), except
that the external force acting on the system is no longer a Gaussian
white noise, but a coloured noise, an Orstein-Uhlenbeck process. The
first merit of this system is that henceforth, there is no possible
confusion in the interpretation of our system: the situation described
by \myref{eqbase} does not correspond anyhow to a thermalization
process. Actually, a Langevin equation describing the thermalization
of a particle in a thermal bath with finite correlation time
(exponentially correlated), would have involved a frictional term
\begin{align}
\int_0^t du (\Ga/k_BT\tz)e^{-(t-u)/\tz}v(u)
\end{align}
instead of our single $\ga v$ \cite{kubo}.

The second merit of the system is to provide us with one dimensionless
quantity $\al=\ga\tau_0$ (in the free case $\om=0$), which is a generalization of the previous
case where only one typical energy can be constructed. It allows to
test the reaction of a given system under sollicitations which differ
only by a characteristic time.

The details of calculations are fully postponed in the first
appendix. Here we discuss only the results obtained. The large
deviation function $f(\eps)$ has again generically two expressions
depending on the general shape of the initial velocity distribution. If this
distribution  is bounded at $t=0$, the large deviation
function has the expression (with the definitions $\be=\om\tau_0$ and $\teps=\eps/\Ga$))
\begin{align}\label{fpremiere}
  f(\eps)=\frac{1}{16\tau_0}\left(-\frac{\teps}{\al}(\al^2+1-2\be^2)\th^2(\teps)-6\frac{\teps}{\al}(\be^2+\frac{\al}{\teps})\th(\teps)+(\al+1)(8+\frac{\teps}{\al}(\al-1)(\al^2-1-4\be^2))\right)\\
\text{where $\th(\eps)$ is the largest real root of $\th^3-(\al^2+1-2\be^2)\th-2(\be^2+\al/\teps)=0$}\nonumber 
\end{align}
An alternative and useful expression of this function is also
\begin{align}
  &f(\eps)=g(\la)+\la\eps\ \ \text{with $\la$ such that
  $g'(\la)=-\eps$}\nonumber\\
&g(\la)=\frac{1}{2\tau_0}\left(\al+1-
\widehat{\th}(\la)\right)\ \ \text{with $\widehat{\th}(\la)$ the
  largest positive root of}\label{fseconde}\\
&(\th^2-\al^2-1+2\be^2)^2-8\be^2\th-4(\be^4+\al^2-2\be^2+4\Ga\la\ga\tau_0^2)=0\nonumber
\end{align}
(remark that $g$ is  the Legendre transformation of
$f$).
 If $\om=0$, the previous expressions are simplified; for example we get 
  \begin{align}
g_{\be=0}(\la)=\left(\al+1-\sqrt{\al^2+1+2\al\sqrt{1+4\Gamma\la/\ga}}\right)/(2\tz),
  \end{align}
 and $f$ could be as well explicitely computed, since the inversion of $g(\la)$ involves roots of a third degree polynomial. Incidentally it can be checked that \myref{fseconde} gives the correct limit for $\tau_0\rightarrow0$: owing to the fact that $\te=1+O(\tau_0)$,  the leading order yields
 \begin{align}
(\te^2-1)^2=8\be^2+4(\al^2-2\be^2+4\Ga\la\ga\tau_0^2)
\end{align}
whence $g_{\tau_0=0}(\la)=\frac{\ga}{2}(1-\sqrt{1+4\Ga\la/\ga})$ whatever the value of $\be$, as established in \cite{jean}.

\begin{figure}[h]
\centerline{\resizebox{14cm}{!}{\includegraphics{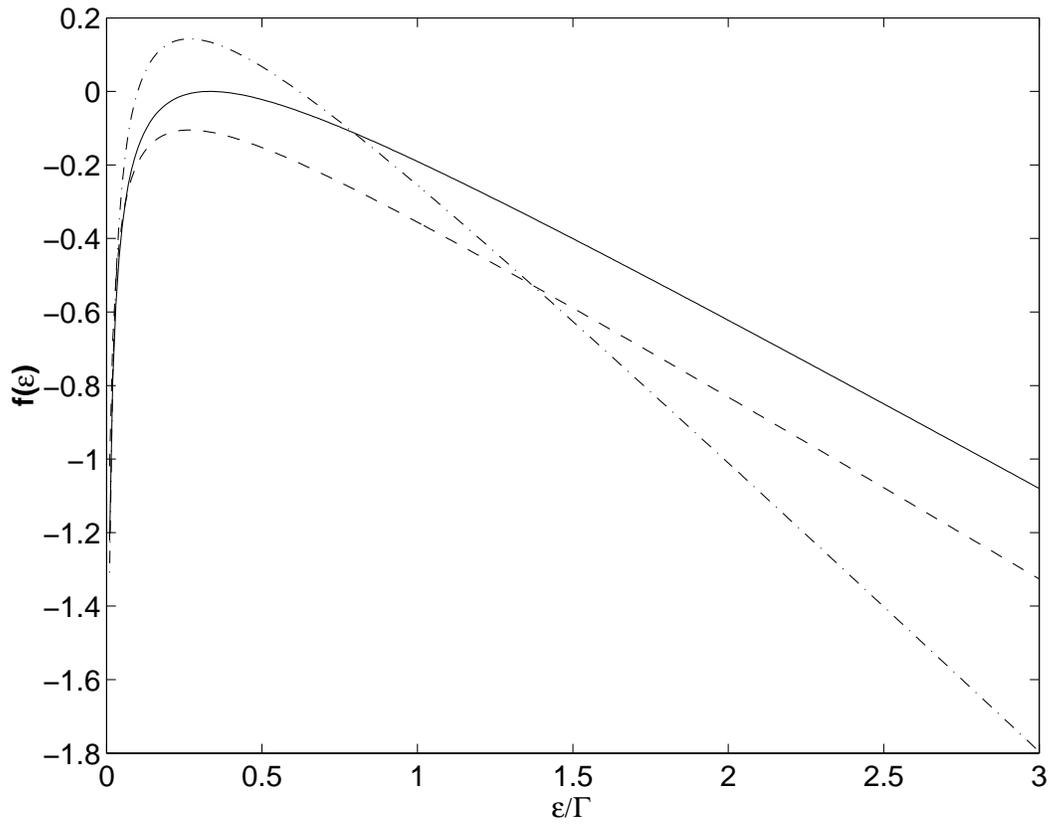}}}
\caption{Typical shape of $f(\teps)$ for $\al=0.5, 1, 2$ and
  $\be^2=(\al^2+1)/2$ (respectively dashed, plain, dash-dotted)}\label{typicalshape}
\end{figure}

The typical shape of that function is given on the figure
\ref{typicalshape}
and is typically asymmetric. The principal characteristics of these
 functions are the location and the curvature of the maximum, and the
 nature of the divergence at $\eps\rightarrow 0$ and $\eps\rightarrow\infty$.
 The maximum of $f$ is obtained for the average injected power $\eps=\overline{\eps}$ (and
correspondingly $\la=\overline{\la}=0$ since
$0=[g'(\overline{\la})+\overline{\eps}]d\la/d\eps+\overline{\la}$),
\begin{align}
\overline{\eps}=\frac{\Ga}{\al+1+\be^2}  
\end{align}
It is interesting to note that the presence of a confining harmonic potential
is always a hindrance to injection of energy into the system. Whether this fact is absolutely general whatever the form of the
(confining) potential is still an open question.

\begin{figure}
  \centerline{\resizebox{14cm}{!}{\includegraphics{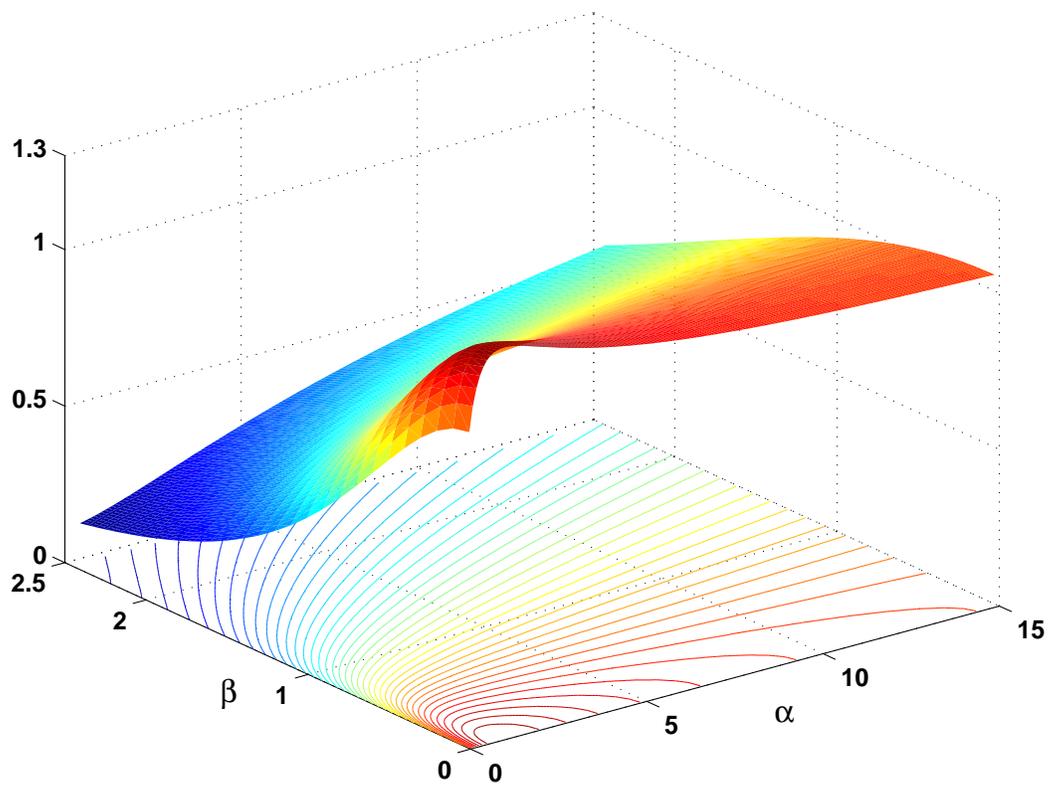}}}
\caption{Surface plot of $T_{curv}(\al,\be)$. We took $\Ga/\ga=1$ for convenience.}\label{Tcurv}
\end{figure}

The curvature of $f$ near the maximum $\overline{\eps}$ can also be
explicitely computed, for it is easy to show that
$f''(\overline{\eps})=-1/g''(0)$. We obtain
\begin{align}
 \sigma_I^2\equiv 1/|2f''(\overline{\eps})|=\frac{\Ga^2}{\ga}\frac{\al^2+3\al+1+\be^2}{(\al+1+\be^2)^3}
\end{align}
This quantity is relevant as one easily measured (numerically or
experimentally) besides the mean value. Incidentally, we can form
with $\bareps$ and $\sigma_I^2$ a characteristic energy
\begin{align}
  T_{curv}\equiv\frac{\sigma_I^2}{\bareps}=-\frac{g''(0)}{2g'(0)}=\frac{\Ga}{\ga}\left(1+\frac{\al-\be^2(1+2\al)-\be^4}{(\al+1+\be^2)^2}\right)
\end{align}
The form of $\frac{\ga}{\Ga}T_{curv}(\al,\be)$ is plotted on figure
\ref{Tcurv}. 
We shall see that this characteristic energy $T_{curv}$ is  in a certain sense an invariant of the
energy flow process (see later).

Let us look now at asymptotic branches. The $\eps\rightarrow 0$ branch
is characterized by a $\eps^{-1/3}$ divergence:
\begin{align}
  f(\eps)\sur{\sim}{\eps\rightarrow 0}\ -\frac{3\ga}{4^{4/3}\al^{2/3}}\left(\frac{\Ga}{\eps}\right)^{1/3}
\end{align}
This exponent $-1/3$ is a novelty with respect to the ``pure'' Langevin
case, where the exponent was $-1$. This is not very intuitive, but
follows closely the decreasing of $\bareps$ with $\al$: it is probably
due to the fact that $I=\psi v$ is bounded as soon as $\al\neq 0$, which
was not the case with a white noise; this important change in the
shape of $f(\eps)$ shows that the white noise limit is a singular
case, even for the function $f(\eps)$ which is a priori related to a
time-integrated observable. Besides, it shows that the short time
dynamical details continue to play a role in the large deviation
function\ldots We can just remark that this vicinity of
$\eps=0$ becomes independent of $\be$: we can thus conjecture that
this exponent is potential independent and is controlled only by the
injection and dissipation mechanisms.

\begin{figure}
  \centerline{\resizebox{14cm}{!}{\includegraphics{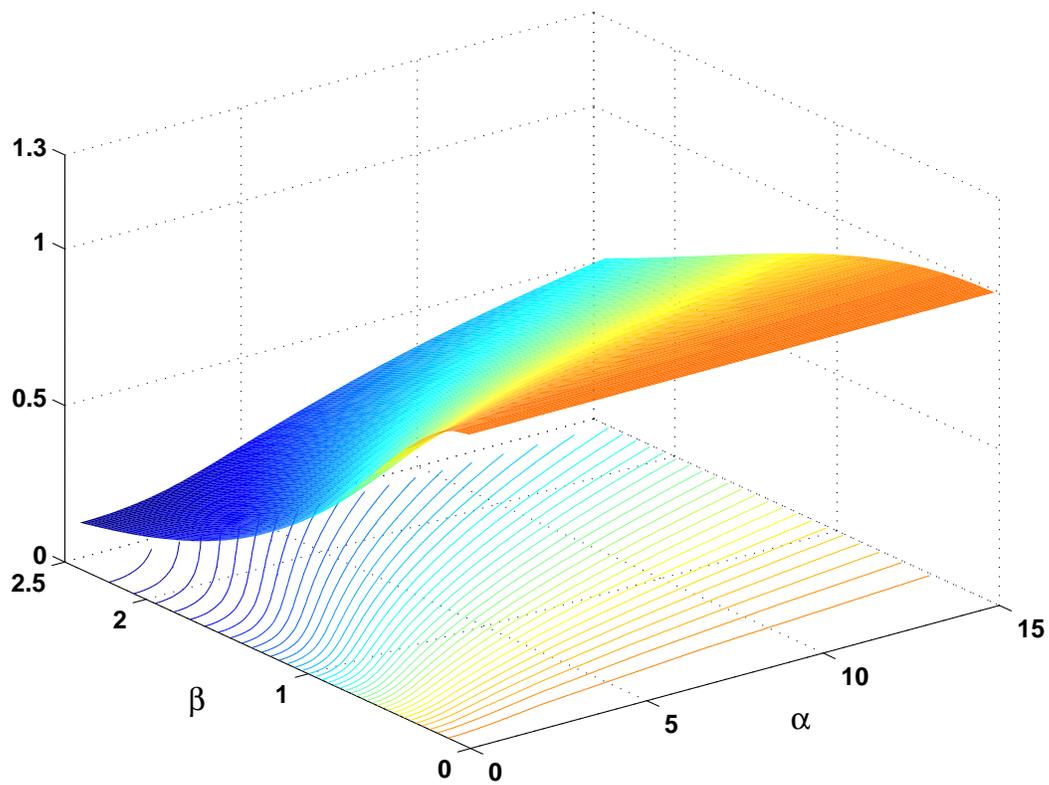}}}
\caption{The energy $T_{slope}$ associated with the
  $\eps\rightarrow\infty$ divergence. We took $\Ga/\ga=1$ for convenience.}\label{Tslope}
\end{figure}

\begin{figure}
  \centerline{\resizebox{14cm}{!}{\includegraphics{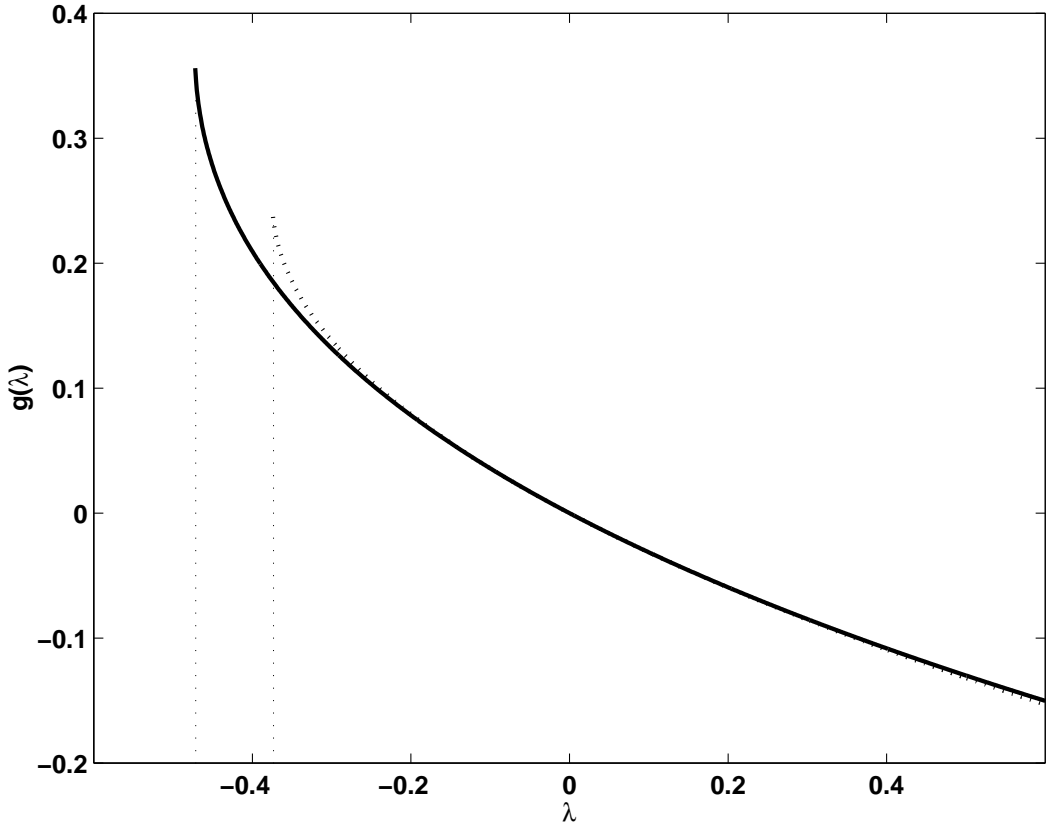}}}
\caption{Exact function $g(\la)$ ($\al=\be=1$) together with the
  parabolic Ansatz (dots) based on $T_{curv}$ (cf. text for details)}\label{typicalg}
\end{figure}

The other asymptotics corresponds to $\eps\rightarrow\infty$, and is much more
softer than the $\eps=0$ one, that is a simple
linear behaviour, witness of a singularity of $g(\la)$ occuring for a
finite negative value $\la=\la_0$. The slope of the line is
homogeneous to $(\text{energy})^{-1}$. In the pure Langevin case, $f(\eps)=-\ga
\Ga((\eps/\Ga)-1)^2/4\eps$ and the slope is simply $-\ga/4\Ga$. We are thus
naturally led to define a characteristic energy $T_{slope}$ by
asserting that the slope of the $\infty$ divergence is given by
$-1/4T_{slope}$.From \myref{fpremiere}, 
\begin{align}
  &T_{slope}=-4\al^2\frac{\Ga}{\ga}\left(-(\al^2+1-2\be^2)\zeta^2-6\zeta\be^2+(\al^2-1)(\al^2-1-4\be^2)\right)^{-1}\\
&\text{where $\zeta$ largest solution of}\ \ \zeta^3-(\al^2+1-2\be^2)\zeta-2\be^2=0
\end{align}
The figure \ref{Tslope} gives a plot of $\frac{\ga}{\Ga}T_{slope}$ as a
function of $\al$ and $\be$. 
The behaviour of $T_{curv}$ and $T_{slope}$ is qualitatively quite
similar, what is not so surprising: the typical form of $g(\la)$ is
shown on figure \ref{typicalg}. If we try for $g$ (in the vicinity
of $\la=0$) the Ansatz
$g(\la)\approx g(\la_0)-\mathrm{C^t}\sqrt{\la-\la_0}$, we deduce
that would
 this Ansatz be exact, we would have $T_{curv}=T_{slope}$.
As
this Ansatz is not so bad (cf. fig. \ref{typicalg}), we deduce
$T_{curv}\approx T_{slope}$. Thus, the large deviation function $f(\eps)$ has its right
asymptotic branch slightly constraint by the vicinity of its maximum,
 but this relationship remains however relatively weak (the relative
 difference does not exceed $45\%$). Physically, the energy $T_{curv}$
 is more interesting, since it is constructed with the top of the
 curve, i.e. with quantities which are easy to measure/compute, which
 is not the case with the asymptotic tails, associated to 
 rare events. 

\subsection{Singularity of $f$ in the permanent regime}

From the preceding paragraph, it is clear that unlike the white noise
case, the large deviation function of $\eps$ is sensitive to the short time dynamics and to
the presence of a potential. Thus, the irrelevance of $V(x)$
demonstrated in \cite{jean} is specific to the white noise case and it is probably hopeless to seek in
general for such simplifications. Nevertheless, something already
observed in \cite{jean} is still present here and is probably
extremely general: in the permanent regime --that is, when the initial
conditions are not fixed, but sampled from the stationary distribution--,
the shape is altered and a negative tail appears below a certain
\textit{positive} value $\eps_c$. This tail is simply the straight
line $g(\la_c)+\la_c\eps$ where $\la_c$ corresponds to $\eps_c$. This
effect corresponds to the fact that for small values of $\eps$, the
probability of these rare events is no longer dominated by the inner
dynamics (inside the time interval $[0, \tau]$) but by rare and very
energetic initial conditions (cf. \cite{jean} for details). This
situation is due to the fact that $I$ and $D$ are intimately coupled
to the energy which has a conservative character. As a result, this
phenomenon must also occur in the models considered here, and more
generally in any case where the stationary distribution of energy is
unbounded. 
As an example, consider the case $\be=0$ (for the sake of simplicity). The
computation of the so-called ``fluctuation term'' in $\lan
e^{-\la\tau\eps}\ran$ provides a prefactor to $\exp(-\tau g(\la))$
which displays a cut for $\la>\la_c=2\ga(2\al^2+3\al+1)/\Ga$.  The corresponding value for $\eps$ is
  \begin{align}
    \eps_c=\frac{\Ga}{\sqrt{9+16\al^2+24\al}}\frac{1}{\left(\al^2+1+2\al\sqrt{9+16\al^2+24\al}\right)^{3/2}}
  \end{align}
and tends to $1/3$ when $\tau_0\rightarrow 0$, as expected from \cite{jean}
This singularity has an interesting physical meaning: it is intimately
related to the three observables $(E,I,D)$ and their mutual dependence
and balance (the stationary distribution is a property of the
dynamics). The disadvantage of that phenomenon is that it is located
in a region of quite rare events, and corresponds to a second order
singularity: as a result, it is not easily observable. One way to
make the singularity observable  could be to consider ``very hot'' and
externally controlled initial
conditions, where large initial energies be likely; this will
certainly shift the location of the singularity. For instance, in the
pure Langevin case $\al=\be=0$ it can be shown that it shifts the
singularity from $\eps/\Ga=1/3$ to $\eps/\Ga=1/\sqrt{3+2\sqrt{2}}\approx
0.414$ (in the limit of flat initial condition). Thus, this shift
would be anyway limited, but could be however sufficient to make the
singularity measurable. In that case, precursors of this singularity
could be observable at finite $\tau$ by plotting $\frac{1}{\tau}
d^2(\log P)/d\eps^2$: this function must display a steep jump at the
singularity location.

\subsection{Characteristic time of the energy flow}

Above was defined a quantity  related to the curvature at the top of
$f(\eps)$ as
$\sigma_I^2=1/|2f''(\bareps)|$. In fact $\sigma_I^2$ is also given by the
following expression:
\begin{align}
\sigma_I^2=\int_0^\infty dt [\lan I(t)I(0)\ran_{st}-\lan I\ran_{st}^2]  
\end{align}
where $\lan\ldots\ran_{st}$ denotes the averaging in the
(out-of-equilibrium) stationary regime.
To demonstrate that, let us compute $\lan \eps^2\ran-\lan\eps\ran^2$
in the large $\tau$ limit. We have
\begin{align}
P(\eps)=p(\eps).\exp(\tau f(\eps)) \sim p(\lan \eps\ran).\exp[\tau
 f''(\lan \eps\ran)(\eps-\lan  \eps\ran)^2/2]
\end{align}
 where $p(\eps)$ is the preexponential factor 
of the distribution. It is readily obtained $\lan
\eps^2\ran-\lan\eps\ran^2\sim 1/|\tau f''(\lan\eps\ran)|$. Expanding
the second moment, the result is established.

In all the stochastic models we considered so far, we always noticed
that large deviations functions associated with energy injection and
energy dissipation were intimately related, and even equal in the
vicinity of the maximum. It is easy and instructive to demonstrate
this property (which is always valid) directly from the expression of
$\sigma_I^2$ (we recall that the stationarity implies $\lan
I\ran_{st}=\lan D\ran_{st}$). Indeed, 
we have
\begin{align}
 \sigma_I^2&= \int_0^\infty dt \left[\lan
 \intvide(\dot{E}(t)+D(t))(\dot{E}(0)+D(0))\ran_{st}-\lan
 D\ran_{st}^2\right]\\
&=\lan(E(\infty)-E(0))(\dot{E}(0)+D(0))\ran_{st}+\int_0^\infty dt \left[\lan
 \intvide D(t)(\dot{E}(0)+D(0))\ran_{st}-\lan
 D\ran_{st}^2\right]\\
&=\lan E\ran\lan D\ran-\lan E(0)D(0)\ran_{st}+\int_0^\infty dt\ \left[\lan
 \intvide D(0)(\dot{E}(-t)+D(t))\ran_{st}-\lan
 D\ran_{st}^2\right]\\
&=\int_0^\infty dt [\lan D(t)D(0)\ran_{st}-\lan D\ran_{st}^2]  =\sigma^2_D
\end{align}
where we made use only of time translation invariance, assumed in a
stationary regime. This property allows us to define some quantities
naturally associated with the energy dynamics. We already saw
$T_{curv}$ which thus can be computed from the fluctuations of the
dissipation term, but the very physical interpretation of this energy
is not clear up to now. Similarly, we 
 define a characteristic
time $\tau_e$ associated with the energy
  dynamics,
\begin{align}\label{dosoldooooooooomire}
  \tau_e=\frac{\sigma_I^2}{\lan I\ran^2_{st}}=\frac{\sigma_D^2}{\lan D\ran^2_{st}}
\end{align}
which presents also the remarkable property to be  symmetric with
respect to the injection and the dissipation. These considerations are
extremely general and make use only of the $\dot{E}=I-D$ structure and
the time translation invariance. Thus \textit{the equation
\myref{dosoldooooooooomire} is always true} in the stationary
regime, and $\tau_e$ is a correlation time attached to the
full energy flow process. 
Usually, we define the correlation time of
an observable as $\tau_X=\left[\int_0^\infty dt \lan
  X(t)X(0)-\lan X\ran^2\ran\right]\left/\lan X^2-\lan
X\ran^2\ran\right.$. Thus, $\tau_e$ is more or less a bare time whence
$\tau_I$ and $\tau_D$ are constructed as
\begin{align}
  \tau_I=\frac{\lan I\ran^2}{\lan I^2\ran-\lan I\ran^2}\tau_e
\end{align}
(and a similar one with $D$). These relations are interesting, since
if we imagine that $\tau_e$ is fixed (or takes it as the natural
time unit), we see that at the injection
level, the fluctuations level is enslaved by the
characteristic time of the injection mechanism. At the other side, this balance
exists with a dissipation time which is possibly quite different from the former; in that case the
fluctuations are again constrained to adjust correspondingly. This simple
reasoning of course does not take into account the fact that the
``reference'' time $\tau_e$ is itself a property of the established regime.

As an example, let us look again at our toy-model. 
Noticing that $\tau_e=T_{curv}/\bareps$, we get
\begin{align}
\tau_e=\frac{1}{\ga}\ \frac{\al^2+3\al+1+\be^2}{\al+1+\be^2}
\end{align}
This result is independent of $\Ga$ the strength of the excitation, a fact
 obvious in our case, since no dimensionless number can be made with
$\Ga$. But for more complicated models a $\Ga$ dependence is to be
expected \textit{a priori}. We notice also
 the effect of the confining potential: the oscillation time $\om$
 only  prevents the correlation time $\tau_e$ to diverge when $\tau_0$
 is large. This is obviously due to the unavoidable oscillation at
 frequencies near $\om$ in the response function of $x$.

\section{Anharmonic potentials}

\begin{figure}[h]
\centerline{\resizebox{14cm}{!}{\includegraphics{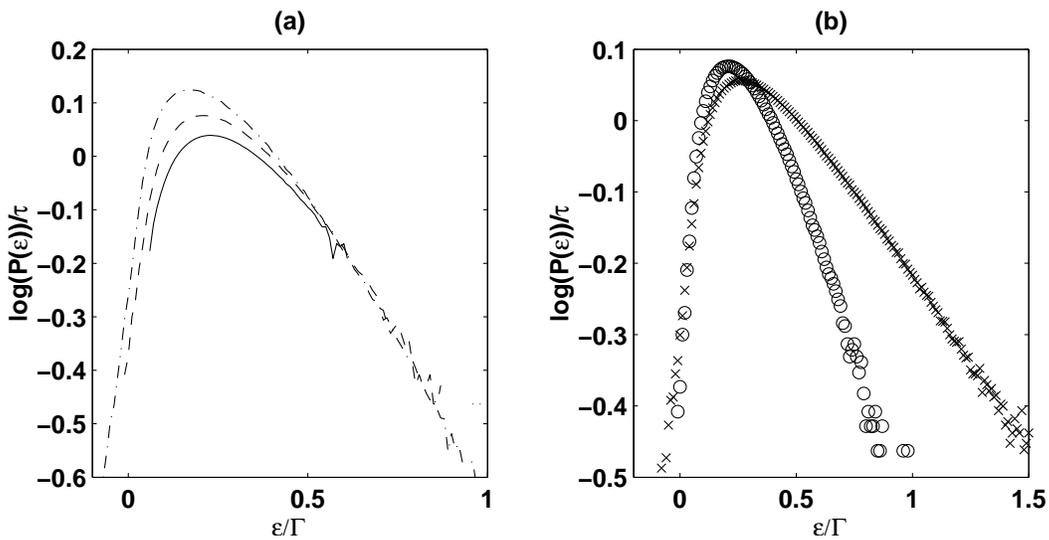}}}
\caption{Typical shape of the distribution $P(\eps)$ for a double well
  confining potential (See details in text). (a): $P(\eps)$ for
  different values of $\tau$ ($\tau=10$, dash-dotted, $\tau=20$,
  dashed, $\tau=50$, plain). (b): for $\tau=20$, comparison between
  double well potential case (circles) and harmonic potential (crosses)}\label{doublewell}  
\end{figure}

Unlike the pure Langevin model, the addition of an anharmonic confining
potential makes the computation of $f(\eps)$ untractable. Actually the
mapping to an effective dynamics (as exposed in  appendix) is no longer feasible,
due to a proliferation of new types of terms in the action. However,
the previous analysis will be qualitatively the same, and the pdf of
$\eps$ has typically the same asymmetric shape, with a right tail
driven by a term $\exp(-\tau \eps/T_{slope})$ (this term is in general
multiplied by a prefactor $\sim \eps^{-\nu}$ where $\nu$ is potential
dependent \cite{jean}): figure \ref{doublewell}(a) shows the quantity
$\frac{1}{\tau}\log P(\eps)$ for a model $\al=\be=1$, $\Ga=1$ and a double well
potential $V(x)=0.25*x^2-x^4$. We see that the convergence
of $\log(P)/\tau$ to $f(\eps)$ is quite slow, due to correction terms
of order $\log(\tau)/\tau$. The other part of the figure shows the
comparison of two pdf for the same value of $(\tau,\al)$, but for two
different potentials (harmonic $\be=1$ and double-well). It clearly shows a
strong dependence of $\log(P)$ (and therefore $f$) with respect to the
potential.
 The positive tail is less pronounced in the nonlinear case,
 which is a bit surprising, since the parameter was chosen such that
 the curvature in the bottom of the double well is identical to the
 harmonic curvature. Naively, as in a double well $\varphi^4$
 potential the frequency decreases with the energy we would expect a
 scenario similar to an effective harmonic potential with a lower well
 frequency, and thus, an enlargement of the rare positive events. This
 is not observed, thus demonstrating that probably rare energetic events correspond to scenarios where the steep external branches of the potential are often explored by the particle -- very energetic oscillations have \textit{in fine} an increasing oscillation.
Finally, a second order singularity in $f$ is also
expected in the permanent regime, although difficult to observe, as
easily seen in figure \ref{doublewell} (a): for $\tau=50$ (whereas the
typical dynamical time are of order $1$), the difference
$\frac{1}{\tau}\log(P)-f$ is not negligible (cf the maximum of the
curve), and several millions of statistic steps are however unable to sample
the vicinity of $\eps=0$\ldots 
To conclude, let us mention that a variational approach could be performed on these nonlinear
 models: using $\lan e^X\ran\geqslant e^{\lan X\ran}$, we could compute for each $\la$ the best harmonic potential describing the dynamics.

\section{Nonlinear friction}\label{nlfriction}

There is another way to generalize the original Langevin equation,
namely to change the dissipation term $\ga v$ to a nonlinear term $
\varphi(v)$, where $\varphi$ is an odd function, positive for positive
$v$. It is worth looking
at these models, since we can that way test the influence of the dissipation
term on the shape of $P(\eps)$. As an example, let us consider the
simplest case
\begin{align}\label{nlfr}
  \dot{v}+\varphi(v)=\psi(t)\ \ \text{with}\ \ \lan\psi(t)\psi(t')\ran=2\Ga\de(t-t')
\end{align}
It is shown in the appendix that the function $g(\la)$ can be
expressed in terms of the lowest energy of a Schr\"odinger equation:
\begin{align}\label{glanl}
  g(\la)/Ga=-\sur{\text{Min}}{\lan\zeta|\zeta\ran=1}\left\lan
  \zeta\left|-\frac{\pa^2}{\pa
  v^2}+\frac{1}{4\Ga^2}\left(\varphi^2(v)-2\Ga\varphi'(v)+4\Ga\la v\varphi(v)\right)\right|\zeta\right\ran
\end{align}
whence we get {\renewcommand{\arraystretch}{0.52}
\begin{align}
  f(\eps)/\Ga=-\sur{\text{Max}}{\begin{array}{c}\scriptstyle \lan\zeta|v\varphi|\zeta\ran=\eps \Ga\\ \scriptstyle \lan\zeta|\zeta\ran=1\end{array}}\left\lan
  \zeta\left|-\frac{\pa^2}{\pa
  v^2}+\frac{1}{4\Ga^2}\left(\varphi^2(v)-2\Ga\varphi'(v)\right)\right|\zeta\right\ran
\end{align}
}
Thus, the large deviation function comes in this case from an
extremalization principle, but this fact is specific to the situation
considered. The inspection of \myref{glanl} is instructive, for we
see immediately that if $\varphi(v)$ diverges faster than $v$,
$g(\la)$ has no longer a singularity at finite $\la$. As a result, the
right asymptotic tail of $f(\eps)$ is no longer a straight line: the
precise form of this asymptotics depends on the dissipation
efficiency. And we find obviously that if
$\varphi(v)/v\rightarrow\infty$, the right tail of $f(\eps)$ goes
faster to $-\infty$. To be more quantitative, if $\varphi(v)\sim
v^{\nu}$ for large $v$ (with $\nu>1$), we can show by considering the effective
potential in \myref{glanl} that $f(\eps)\sim -\eps^{2\nu/(\nu+1)}$.
Does it affect the left asymptotics as well ? To answer this question,
 the $\la\rightarrow +\infty$ limit must be investigated. The effective potential
is deeply changed by this change of sign and is now single well and is
equivalent to $\la v\varphi(v)$. As a result we expect a minimum
energy scaling like $\la^{1/2}$ \cite{cct} (provided $\varphi(v)\sim v$ near
zero). This leads to a divergence $\eps^{-1}$ near zero, whatever the
precise form of $\varphi(v)$. However the condition $\varphi\propto v$
near $v=0$ is crucial, otherwise the left tail of $f$ is affected: if
for instance $\varphi\sim v^3$, the zero point energy of the
Schr\"odinger potential scales like $\la^{1/3}$ and we find $f(\eps)\sur{\sim}{0}\eps^{-1/2}$.

This example is quite interesting, since it shows that the dissipation
mechanism can affect deeply the general shape of $f(\eps)$, and
particularly the form of the asymptotic tails. In particular, the
behaviour of $\varphi(v)$ for small velocities dictates the form of
$f(\eps)$ near $\eps=0$, whereas correspondingly the large $v$
behaviour of $\varphi$ is responsible for the large $\eps$
value. Thus, for realistic situations, a study of the asymptotic tails of the $\eps$ distribution could provide an estimation of an effective (in average) dissipation/energy relation: in a turbulent flow for instance, the greater the energy present in the system, the greater the dissipation; we argue that the implicit relation between these observables is partly encoded in the asymptotic tails of $f(\eps)$.

\section{Conclusion}

In this paper, we studied the injected power distributions of several
stochastic models, in order to precise the results obtained in
\cite{jean} and test their possible generality. This study showed
that, in general, the large deviation functions remain sensitive to
the details of the short-time dynamics of the system, despite the time
averaging: this observation makes the Langevin-like models very
peculiar, since in these cases the large deviation function is
insensitive to the presence of a pinning potential. 
However, the common scenario of dissipative systems in a permanent
regime makes all these  to share some common characteristics:
first, they are strongly asymmetric, which reflects the fact that we
follow a positive conserved quantity, with a systematic dissipation;
secondly, they display a second order singularity in the permanent
regime which constructs a negative tail in the large deviation
function associated to the injected power (this singularity is not
present in the dissipation, since the dissipation is always
positive). Third, the large deviation functions of injected and dissipated power have the
same curvature at their maximum, due to the structure of the energy
equation. We can even conjecture that in general the two large deviation functions
are strictly equal in a vicinity of the maximum (it is easy to show it
for the models presented here, but probably often true). From this
equality  a characteristic time (or a characteristic
energy) naturally associated to the energy flow can be defined. It would be interesting
to measure that time in an extended dissipative system, when the
system is chaotic and generates itself its disorder and the noisy
character of the energy injection (instead of being imposed by the
operator): the typical time scale is in that case  controlled by
the dynamics of the system itself (for instance, what is the behaviour
of $\tau_e$ near an instability threshold ?) .   More generally, it would be worth
now  performing similar analysis to extended systems where non
equilibrium stationary states can produce structurations of the system.

\section{Appendix A: large deviation function for the non markovian system}

We derive in this appendix the main results of section \ref{coloured}. We consider  the injected power averaged over a time
interval $\tau$ :
\begin{align}
\eps&=\frac{1}{\tau}\int_0^\tau dt \psi(t)v(t)
\end{align}
for the model \myref{eqbase}.
As $\tau$ is finite, this observable remains a fluctuating quantity,
and we examine  its probability distribution function  $\pi(\eps)$. More precisely, we are interested in the computation of the so-called \textit{large deviation function}, a function which emerges from the consideration of large values of $\tau$. It can be proved that for large values of $\tau$ the probability of $\eps$ verifies
\begin{align}
\log\pi(\eps)\sur{\sim}{\tau\rightarrow\infty}\tau f(\eps)
\end{align}
(this equivalence is sometimes boldly noted $\pi(\eps)\sim e^{\tau
f(\eps)}$, understanding the prefactor). The function $f$ is nothing
but the large deviation function associated with $\eps$. We
concentrate ourself on this sole quantity, since the prefactor is
quite more involved to evaluate, and is physically much less
interesting.

For convenience, we are going to compute two different types of pdf,
namely $\pi_\wz(\eps)$ the probability of $\eps$, \textit{knowing that
  the initial conditions of the process are
  $\wz=(x(0)=x_0,v(0)=v_0,\psi(0)=\psi_0)$}, and $\pi(\eps)$ the probability of
$\eps$ in the permanent (stationary) regime. The latter is simply
related to the former through
\begin{align}
\pi(\eps)=\int dx_0dv_0d\psi_0\ P_{st}(x_0,v_0,\psi_0)\pi_\wz(\eps)
\end{align}
where $P_{st}$ is the stationary distribution of the (correlated)
variables $(x,v,\psi)$. As $\pi$ and $\pi_\wz$ are different, their large deviation functions  are termed $f_\wz$ and $f$ respectively henceforth.

\subsection{Path integral representation of the characteristic function}

The characteristic function of $\pi_\wz(\eps)$ is defined by
\begin{align}\label{piz}
\piz&=\int d\eps \ \pi_\wz(\eps) \exp(-\la\tau\eps)
\end{align}
As soon as this function is known, the original pdf  can be retrieved via an inverse Fourier transform (in the complex $\la$ variable). At the level of the large deviation function, equation \myref{piz} gives a rapid answer: there exists a function $g_\wz(\la)$ such that
\begin{align}
\piz\sur{\sim}{\tau\rightarrow\infty} \exp[\tau g_\wz(\la)]
\end{align}
(in the bold sense !), and this function is given by the Legendre transform of $f_\wz$:
\begin{align}
g_\wz(\la)&=f_\wz(\eps)-\la\eps\\
f_\wz'(\eps)&=\la
\end{align}
(the inversion is simply: $f_\wz=g_\wz+\la\eps, g_\wz'(\la)=-\eps$). All that is a priori correct, but for a point: it is possible that the prefactor of $e^{\tau g}$ in the characteristic function diverges for a particular value of $\la$, and that therefore the function be not defined for certain values of $\la$, \textit{whereas the function $g$ remains defined in this range}. In that case, the rapid Legendre inversion above mentioned must be replaced by a careful analysis of the integral \myref{piz} (see \cite{jean} and below).

\bigskip

Let us give now a path integral representation of $\piz$. It is quite easy to do that in our case, since the noise $\psi$ is nothing but a Ornstein-Uhlenbeck, that is a process described by an ordinary Langevin equation:
\begin{align}
\dot{\psi}+\tz^{-1}\psi&=\zeta(t)\\
\langle \zeta(t)\zeta(t')\rangle&=2\frac{\Gamma}{\tz^2}\de(t-t')
\end{align}
where $\zeta$ is a Gaussian white noise. The path-integral
representation of the propagator of this process is known
\cite{jean,wiegel,feynman,zj} :
\begin{align}
P(\psi_1,\tau|\psi_0,0)=e^{\frac{\tau}{2\tau_{ 0}}}\times\int_{\psi(0)=\psi_0}^{\psi(\tau)=\psi_1}[\mathcal{D}\psi] \exp\left(-\frac{1}{4\Gamma\tz^{-2}}\int_0^{\tau}dt\left(\dot{\psi}+\tz^{-1}\psi\right)^2\right)
\end{align}
As a result, the statistical weight of a particular occurence of the
noise, knowing that it originates from $\psi_0$ at $t=0$ is given by
\begin{align}
e^{\frac{\tau}{2\tau_{ 0}}}\times\exp\left(-\frac{1}{4\Gamma}\int_0^{\tau}dt\left(\tz\dot{\psi}+\psi\right)^2\right)
\end{align}
Therefore, we get the probability of a particular occurence of the process $x$ (beginning at $\wz$) as
\begin{multline}
Prob\left[\intvide[x(t)],t\in[0,\tau]\left|\intvide\right.x(0)=x_0,v(0)=v_0,\dot{v}(0)=\psi_0-\ga v_0-V'(x_0)\right]\\
=e^{\frac{\tau}{2\tau_{ 0}}(1+\al)}\times\exp\left(-\frac{1}{4\Gamma}\int_0^{\tau}dt\left[\tz(\ddot{v}+\ga\dot{v}+vV''(x))+(\dot{v}+\ga v+V'(x))\right]^2\right)
\end{multline}
where we defined $\al=\ga\tz$, and $V(x)=\om^2x^2/2$.
It must be noted that the jacobian of the transformation $\psi\rightarrow x$ is equal to $\exp(\ga\tau/2)$ and is taken into account in the preceding expression.

From now on, it is easy to deduce an expression for $\langle e^{-\la\tau\eps}\rangle_\wz$, since the distribution of individual paths is at hand:
\begin{multline}
\left\langle e^{-\la\tau\eps}\right\rangle_\wz \!\!\!=e^{\frac{\tau}{2\tau_{ 0}}(1+\al)}\\
\times \int_{\mathbf{w}(0)=\wz}\!\!\!\!\!\!\![\mathcal{D}x]\exp\left(-\frac{1}{4\Gamma}\int_0^{\tau}\!\!\!dt\left[\tz(\ddot{v}+\ga\dot{v}+vV'')+(\dot{v}+\ga v+V')\right]^2-\la\int_0^\tau \!\!\!dt\ v(\dot{v}+\ga v+V')\right)
\end{multline}
(with obviously $v(t)=\dot{x}(t)$). It is convenient to express explicitely the final values of $\psi$ and $(x,v)$:
\begin{multline}
\left\langle e^{-\la\tau\eps}\right\rangle_\wz \!\!\!=\int d\mathbf{w_1}\ \ e^{\frac{\tau}{2\tau_{ 0}}(1+\al)-\la(E_1^2-E_0^2)}\\
\times \int_{\mathbf{w}(0)=\wz}^{\mathbf{w}(\tau)=\mathbf{w_1}}\!\!\!\!\!\!\![\mathcal{D}v]\exp\left(-\frac{1}{4\Gamma}\int_0^{\tau}\!\!\!dt\left[\tz(\ddot{v}+\ga\dot{v}+vV''(x))+(\dot{v}+\ga v+V'(x))\right]^2-\la\ga\int_0^\tau \!\!\!dt\ v^2\right)\label{mozart}
\end{multline}
where $E\equiv \demi v^2+V(x)$.
The essential point of the derivation is now the possibility to find four constants $U,\widehat{\ga},\widehat{\tau}_0,\widehat{\om}$, such that the difference
\begin{multline}
\Delta=\int_0^{\tau}\!\!\!dt\left([\tz(\ddot{v}+\ga\dot{v}+\om^2v)+(\dot{v}+\ga v+\om^2x)]^2+4\Gamma\la\ga v^2\right)\\
-U\int_0^{\tau}\!\!\!dt\left[\htz(\ddot{v}+\hga\dot{v}+\ho^2v)+(\dot{v}+\hga v+\ho^2x)\right]^2
\end{multline}
implies only boundary terms. If we choose ($U,\widehat{\ga},\widehat{\tau}_0,\ho$) fulfilling (we define of course $\hal\equiv\htz\hga$, and $\be=\tz\om,\hbe=\htz\ho$)
\begin{align}
\tz^2&=U \htz^2\\
\al^2+1-2\be^2&=U(\hal^2+1-2\hbe^2)\\
\frac{1}{\tau_0^2}(\be^4+\al^2-2\be^2)+4\Gamma\la\ga&=\frac{U}{\htz^2}(\hbe^4+\hal^2-2\hbe^2)\\
\frac{\be^4}{\tz^4}&=U\frac{\hbe^4}{\htz^4},
\end{align}
it completes the goal and we can write
\begin{multline}
\left\langle e^{-\la\tau\eps}\right\rangle_\wz \!\!\!=\int d\mathbf{w_1}\ \ e^{\frac{\tau}{2\tau_{ 0}}(1+\al)-\frac{\la}{2}(E_1^2-E_0^2)-\frac{\Delta}{4\Gamma}}\\
\times \int_{\mathbf{w}(0)=\wz}^{\mathbf{w}(\tau)=\mathbf{w_1}}\!\!\!\!\!\!\![\mathcal{D}x]\exp\left(-\frac{U}{4\Gamma}\int_0^{\tau}\!\!\!dt\left[\htz(\ddot{v}+\hga\dot{v}+\ho^2v)+(\dot{v}+\hga v+\ho^2x)\right]^2\right)
\end{multline}
The very reason we made this transformation is that the remaining path integral is closely related to the propagator of a ``renormalized'' process of type \myref{eqbase}, where $(\Gamma,\tz,\ga,\om)\rightarrow (\Gamma/U,\htz,\hga,\ho)$:
\begin{align}
 \int_{\mathbf{w}(0)=\wz}^{\mathbf{w}(\tau)=\mathbf{w_1}}\!\!\!\!\!\!\![\mathcal{D}x]\exp\left(-\frac{U}{4\Gamma}\int_0^{\tau}\!\!\!dt\left[\htz(\ddot{v}+\hga\dot{v}+\ho^2v)+(\dot{v}+\hga v+\ho^2x)\right]^2\right)=e^{-\frac{\tau}{2\tz}(1+\hal)}\times \widehat{P}(\mathbf{w}_1,\tau|\wz,0)
\end{align}
As a result,
\begin{align}
\left\langle e^{-\la\tau\eps}\right\rangle_\wz \!\!\!=\exp\left(\tau\left[\frac{1+\al}{2\tau_0}-\frac{1+\hal}{2\htz}\right]\right)\times \int d\mathbf{w_1}\ \ e^{-\frac{\la}{2}(E_1^2-E_0^2)-\frac{\Delta}{4\Gamma}}\times \widehat{P}(\mathbf{w}_1,\tau|\wz,0)
\end{align}
The remaining integral has a nice behaviour for large $\tau$, since $\widehat{P}(\mathbf{w}_1,\tau|\wz,0)$ tends to its equilibrium value $\widehat{P}_{st}(\mathbf{w}_1)$. Thus, we get the veritable equivalence
\begin{align}
\left\langle e^{-\la\tau\eps}\right\rangle_\wz \!\!\!\sim\exp\left(\tau\left[\frac{1+\al}{2\tau_0}-\frac{1+\hal}{2\htz}\right]\right)\times \int d\mathbf{w_1}\ \ e^{-\frac{\la}{2}(E_1^2-E_0^2)-\frac{\Delta}{4\Gamma}}\times \widehat{P}_{st}(\mathbf{w}_1)
\end{align}
whence we extract  the function $g_\wz$ 
\begin{align}
g_\wz(\la)=\frac{1}{2}\left(\frac{1+\al}{2\tau_0}-\frac{1+\hal}{2\htz}\right)
\end{align}
It is worth noticing that the subscript $\wz$ is useless, for the function $g_\wz$ is independent of $\wz$; we abandon henceforth it (for $g$) in the following.

Let us define $\hat{\te}=(1+\hal)\tz/\htz$. We can verify that $\hat{\te}$ is the largest positive root of
\begin{align}
({\te}^2-\al^2-1+2\be^2)^2-8\be^2{\te}-4(\be^4+\al^2-2\be^2+4\Gamma\la\ga\tz^2)=0
\end{align}
when it exists; if not, $\lan e^{-\la\tau\eps}\ran$ diverges. It proves the result \myref{fseconde}. 
\medskip

Similarly, We get the characteristic function of the process in the permanent regime by simply integrating over $\wz$:
\begin{align}
\left\langle e^{-\la\tau\eps}\right\rangle \sim\exp\left(\tau g(\la)\right)\times \int d\mathbf{w_1}d\wz \ \ e^{-\frac{\la}{2}(E_1^2-E_0^2)-\frac{\Delta}{4\Gamma}}\times \widehat{P}_{st}(\mathbf{w}_1)P_{st}(\mathbf{w}_0)
\end{align}
(Note that the stationary distributions $P$ are not the same for $\wz$ and $\mathbf{w_1}$, since $\widehat{P}$ refers to the renormalized process). We have to remark that the same function $g$ is also shared by the stationary process. But  it does \textit{not} imply that the associated $f_\wz$ and $f$ functions will be the same.

\subsection{Analytical properties of characteristic functions}

To derive the large deviation functions $f$, it is required to look the analytical properties of the prefactor, since the latter can lessen the analyticity domain  of the characteristic function with respect to that implied by the sole inspection of $g$.

A priori, the function $g$ is analytical over the whole complex $\tla$
space, except on a cut located on  (but in general not equal to) $\mathbb{R}^-$. This cut begins when the root $\widehat{\te}$ disappears. In this paragraph, we considers only the free case $\om=0$, for the sake of simplicity. But the discussion remains valid for the bounded case as well.

Let us first consider the non stationary situation:
 here, for a fixed value of $\wz$, the prefactor is composed by
functions as regular as $g$ itself ($g$ has a cut on $\la<\la_-=-\ga/4\Gamma$) times an integral over
$\mathbf{w}_1$. Thus, limitations to analyticity could arise if for
some values of $\la$, this integral no longer converges. A lengthy
computation gives this integral (when converging) proportional to 
\begin{align}
\left[([\sqrt{1+\al^2+2\al\eta}+\al]^2-1)\sqrt{\sqrt{1+\al^2+2\al\eta}-\al+1)}\right]^{-1}
\end{align}
(with $\eta=\sqrt{1+4\Gamma\la/\ga}$) which again is as regular as $g$ itself. It is concluded that for the
non stationary case, the prefactor cannot hinder the 
Legendre inversion, and that $f_\wz$ is really given by the inverse
Legendre of $g$.

Quite different is the stationary case, since now two integrals have
to converge. Of that over $\mathbf{w_1}$ we already proved the
analyticity over $\Cset\setminus \Rset^-$. That implying $\wz$ is
completely different, and  another lengthy computation gives it
proportional to
\begin{align}
(\sqrt{1+\al^2+2\al\eta}+\al+1)^{-1}\left((\eta+1)(3\al+1-\sqrt{1+\al^2+2\al\eta})\right)^{-1/2}
\end{align}
This time, a new cut appears, since for
$\la>\la_c=2(2\al^2+3\al+1)\ga/\Gamma$, the preceding expression is not
defined, which expresses the fact that the integral over $\wz$ is not
defined. This extra cut has deep consequences on the large deviation
function $f$.

\subsection{Large deviation functions}

For the nonstationary case, the usual rule applies, that is $f$ is the
Legendre transform of $g$:
\begin{align}\begin{split}
  f_\wz(\eps)&=g(\la)+\la\eps\\
g'(\la)&=-\eps
	     \end{split}\label{eheh}
\end{align}
For the stationary case, this expression is not always valid:
\myref{eheh} must give $\la<\la_c$ otherwise
there is a problem. In that case, the inverse Laplace transform $\la\rightarrow\eps$ must be lead carefully \cite{jean} but the result is quite simple: if no allowed $\la$ fulfills
\myref{eheh}, the large deviation function for the injection considered is simply
$f(\eps)=g(\la_c)+\la_c\eps$, therefore a portion of a straight line.
Thus, we have
\begin{align}\begin{split}
  f(\eps)=f_\wz(\eps)&\ \ \text{if $\eps>\eps_c$}\\
f(\eps)=g(\la_c)+\la_c\eps&\ \ \text{if $\eps<\eps_c$}
	     \end{split}
\end{align}
and $\eps_c$ defined as $g'(\la_c)=-\eps_c$.

\subsection{Nonlinear potentials}

For a general potential $V(x)$, the expression \myref{mozart} remains true.
 If $V$ is not harmonic, the renormalization procedure is no more possible, since the action involves, besides boundary terms, the terms
\begin{align}
\frac{1}{4\Gamma}\left(\tz^2\ddot{v}^2-\dot{v}^2(\tz^2V''-\al^2-1)+V'^2+v^2(\tz^2V''^2+\ga^2-V'')-\frac{\ga\tz^2}{2}v^3V'''+\frac{\tz^2v^4}{3}V^{(4)}\right)
\end{align}
Thus, there is a ``proliferation'' of new terms which cannot be balanced by a simple redefinition of the constants. In fact, it seems that the integrability of the problem is restricted to the linear case.

\section{Appendix B: nonlinear friction}

Here we sketch the computations pertaining to the section \ref{nlfriction}.
From the dynamical equation \myref{nlfr},
\begin{align}
   \left\lan e^{-\la\tau\eps}\right\ran&=\int[
   \mathcal{D}\psi]\exp\left(-\frac{1}{4\Gamma}\int_0^\tau(\dot{v}+\varphi(v))^2-\la\int_0^\tau
   (\dot{v}+\varphi(v)) v\right)
\end{align}
This integral is a sum over different realizations of $\psi$. To write
it as an integral over realizations of $v$, care must be taken of
the fact that the Jacobian of the transformation is $\demi\int_0^\tau
\varphi'(v)dt$. Then 
\begin{align}
   \left\lan e^{-\la\tau\eps}\right\ran_{v_0}=\int
   dv_1\ e^{-\frac{\la}{2}(v_1^2-v_0^2)}\int_{v(0)=v_0}^{v(\tau)=v_1}[\mathcal{D}v]\
   \exp\left(-\frac{1}{4\Gamma}\int_0^\tau\left[(\dot{v}+\varphi(v))^2-2\Gamma\varphi'(v)+4\Gamma\la v\varphi(v)\right]\right)
\end{align}
The idea is the same as for the linear friction cases: we seek for a
function $\xi(v)$ and a constant $Q$ such that for all $v$
\begin{align}
  \varphi(v)^2-2\Gamma\varphi'(v)+4\Gamma\la v \varphi(v)=\xi^2(v)-2\Gamma\xi'(v)+Q
\end{align}
with the ``initial condition'' $\xi(0)=0$. This is a Riccatti
equation, and the standard transformation
$\xi(v)=-2\Gamma\zeta'(v)/\zeta(v)$ leads to a Schr\"odinger equation:
\begin{align}
  -4\Gamma^2\zeta''+[\varphi^2-2\Gamma\varphi'+4\Gamma\la v \varphi]\zeta-Q\zeta=0
\end{align}
Owing to the fact that the potential $\varphi^2-2\Gamma\varphi'+4\Gamma\la v
\varphi$ is even, that $\zeta$ never vanishes (since $\xi$ must be
defined $\forall v$), we deduce that $Q$ is necessarily the lowest
eigenvalue of the Schr\"odinger problem and $\zeta$ the associated
eigenfunction.

Following the trick of the previous Appendix, we easily deduce that
$g(\la)=-Q(\la)/4\Gamma$, that is, using variational properties of
Schr\"odinger eigenvalues:
\begin{align}
  g(\la)/\Gamma&=-\sur{\text{Min}}{\int
  \zeta^2=1}\int\left[\zeta'^2+\frac{1}{4\Gamma^2}(\varphi^2-2\Gamma\varphi'+4\Gamma\la
  v\varphi)\zeta^2\right]dv\\
&=-\sur{\text{Min}}{\lan
  \zeta|\zeta\ran=1}\left\lan\zeta\left|-\frac{\pa^2}{\pa v^2}+\frac{1}{4\Gamma^2}(\varphi^2-2\Gamma\varphi'+4\Gamma\la
  v\varphi)\right|\zeta\right\ran
\end{align}
To get $f$, we must take the Legendre inversion of this formula:
$f(\eps)=\sur{\text{min}}{\la}[g(\la)+\la\eps]$. As $g'(\la)=-\eps$,
a simple transformation yields:
\begin{align}
  f(\eps)/\Gamma&=-\sur{\text{Max}}{\lan \zeta|v\varphi|\zeta\ran/\lan \zeta|\zeta\ran=\eps}\left\lan\zeta\left|-\frac{\pa^2}{\pa v^2}+\frac{1}{4\Gamma^2}(\varphi^2-2\Gamma\varphi')\right|\zeta\right\ran
\end{align}

\newpage
\pagestyle{empty}

\newpage

\newpage

\newpage

\newpage

\end{document}